\documentclass[modern]{aastex61}
\usepackage{graphicx}
\usepackage{amsmath}
\usepackage{amssymb}
\usepackage{enumitem}

\newcommand\mybar{\kern1pt\rule[-\dp\strutbox]{.8pt}{\baselineskip}\kern1pt}

\setlist[itemize]{noitemsep, topsep=0pt, leftmargin=*}

\shorttitle{Broken Dyson Spheres}
\shortauthors{Loeb}

\begin{document}

\title{Interstellar Objects from Broken Dyson Spheres}

\author{Abraham Loeb}
\affiliation{Astronomy Department, Harvard University, 60 Garden
  St., Cambridge, MA 02138, USA}

\begin{abstract}
Without extensive maintenance, Dyson spheres inevitably
disintegrate by asteroid impacts over billions of years. The resulting
fragments would appear as anomalous interstellar objects, potentially
sharing the unusual shape and motion of 1I/`Oumuamua or the unusual
material strength of the first two interstellar meteors, IM1 and IM2.
If the Dyson sphere's tiles are light sails, the number of fragments
could exceed that of interstellar asteroids because of their reduced
escape speed from the star and the increase in stellar luminosity 
during the red giant phase.

\end{abstract}

\section{Introduction}

%
%
%

In 1937, Olaf Stapledon published the novel ``Star Maker'', in which
he imagined the use of a technologically-manufactored shell of matter
to tap the energy output of a host star. The concept was subsequently
formalized by Freeman Dyson~\citep{1960Sci...131.1667D}, who reasoned
that as the energy needs of humanity will steadily increase, our
civilization might aspire to tap all the energy output of the
Sun. Dyson proposed a shell of orbiting structures that would
intercept and collect the solar luminosity. This so-called Dyson
sphere would emit infrared radiation to balance the heat deposited on
it by sunlight. 
The resulting infrared signal could flag a Dyson sphere in contrast to the
natural optical emission by Sun-like stars. So far, searches for
related infrared signatures from stars or galaxies did not find
evidence for Dyson spheres but only for emission by natural dust (see
references in \citet{2020SerAJ.200....1W}).

Below I demonstrate that if Dyson spheres existed to serve their
civilizations for a limited time, most of them would have
disintegrated within billions of years in the absence of extensive
maintenance. In that case, their ejected fragments could appear as
interstellar objects~\citep{2022AsBio..22.1459S}.

\section{Broken Dyson Spheres}

Rigid Dyson spheres are not easy to maintain. In compliance with Newton's
iron sphere theorem, a perfectly spherical and rigid shell is not
subjected to any net gravitational force from a star interior to it,
regardless of whether it is centered on the star. However, the shell
experiences destructive differential forces across its surface, and
its material strength must be high in order to prevent
deformation. The required tensile strength, $\Pi$, is given by:
\begin{equation}
\Pi \approx {GM_\star \rho_{ds}\over 2R_{ds}}=3.6\times 10^{6}~{\rm
  MPa} \left({M_\star\over M_\odot}\right)\left({\rho_{ds}\over 7.9~{\rm
  g~cm^{-3}}}\right)\left({R_{ds}\over 1~{\rm au}}\right)^{-1} ,
\label{five}
\end{equation}
where $M_\star$ is the stellar mass,
$R_{ds}$ is the associated Dyson
sphere radius and $\rho_{\rm ds}$ is normalized by the solid density of iron.  The
derived value is an order of magnitude above the tensile strength of
graphene, $\Pi_{gr}=1.3\times 10^{5}~{\rm MPa}$.

The required mass of a Dyson sphere with a thickness $W_{ds}$,
\begin{equation}
M_{ds}=4\pi R_{ds}^2W_{ds}\rho_{ds}= 2.3M_{\earth}\left({R_{ds}\over 1~{\rm au}}\right)^2\left({\rho_{ds}\over 7.9~{\rm
  g~cm^{-3}}}\right)\left({W_{ds}\over 1~{\rm
  cm}}\right),
\label{six}
\end{equation}
exceeds the mass of the asteroid belt, $2.4\times 10^{24}~{\rm g}$, by
four orders of magnitude. This may explain the inferred discrepancy
between the required mass in interstellar objects, such as
1I/`Oumuamua, and the expected mass in asteroids ejected from
planetary
systems~\citep{2009ApJ...704..733M,2018ApJ...866..131M,2019AJ....157...86M}.

To circumvent the material strength and mass challenges, Robert Forward proposed
a tiled structure~\citep{1991JSpRo..28..606F}, with each component
functioning as a solar sail for which the star's gravity is exactly
balanced by by the star's outward radiative push, thus maintaining a fixed
position without orbiting the star. The force balance
reads~\citep{2022RNAAS...6..104L}:
\begin{equation}
{L_\star \over 2\pi R_{ds}^2 c}={GM_\star \Sigma_{ds}\over R_{ds}^2},
\label{seven}
\end{equation}
where $L_\star$ is the stellar luminosity and the mass per unit area
of the light sail components is given by,
\begin{equation}
\Sigma_{ds}=\rho_{ds}W_{ds}=1.5\times 10^{-4}~{\rm g~cm^{-2}}\left({L\over
  L_\odot}\right)\left({M_\star\over M_\odot}\right)^{-1}~.
\label{eight}
\end{equation}
In that case, the required thinness of the Dyson sphere sails could
explain~\citep{2022AsBio..22.1392L,2018ApJ...868L...1B} the flat
geometry~\citep{2019MNRAS.489.3003M} and excess push of the
interstellar object 1I/`Oumuamua away from the
Sun~\citep{2018Natur.559..223M} without a cometary
tail~\citep{2018AJ....156..261T}. 

The repulsive push of the stellar radiation pressure is equivalent to
a reduction in the effective gravitating attraction of the star
because both gravitational and radiative accelerations decline
inversely with distance squared. For a ratio $f_{ds}<1$ between the
radiative and gravitational accelerations (the ration between the left
and right sides of equation \ref{seven}), the effective gravitating
mass of the star is $(1-f_{ds})M_\star$.  When the two accelerations
balance ($f_{ds}=1$), as considered by
equations~(\ref{seven}-\ref{eight}), the sail hovers without any
gravitational binding to the star. In that case, the escape speed of
the sail from the star vanishes, and a small gravitational push
outwards by a passing planet could eject it from the planetary system.
Consequently, sail fragments would escape much more easily from a
planetary system, compared to gravitationally bound asteroids which
are trapped by the barrier of a substantial escape speed.

Since sail fragments are thin, they carry much less mass than asteroids of the 
same length, thus relieving the mass discrepancy associated with interstellar 
asteroids~\citep{2009ApJ...704..733M,2018ApJ...866..131M,2019AJ....157...86M}.

Both radiative and gravitational forces are spherically symmetric as
long as the sail maintains a fixed cross-sectional area in the direction of the
star. This can be trivially accomplished by a sail in the shape of a
small sphere which maintains a constant cross-sectional 
area irrespective of orientation.   

Even without damage from asteroids, abandoned sails will eventually be
pushed out of the planetary system as a sun-like star would evolve
through its red giant phase, resulting in an increase in its
luminosity-to-mass ratio.

\section{Implications}

If a Dyson sphere disintegrated by asteroid impacts over time, then its
fragments would have been ejected by a gravitational or radiative push
out of the host planetary system more easily than asteroids.

Interstellar fragments of a broken Dyson sphere could potentially
share the unusual shape~\citep{2019MNRAS.489.3003M} and light sail
characteristics of the interstellar object
1I/`Oumuamua~\citep{2022AsBio..22.1392L,2018ApJ...868L...1B}, or the
unusually high yield strength ($\ga 10^2~{\rm Mpa}$, larger than the
value of iron meteorites) exhibited by the first and second
interstellar meteors, IM1 and IM2~\citep{2022ApJ...941L..28S}. The
planned expedition to retrieve the fragments of
IM1~\citep{2022arXiv220800092S} and the subsequent analysis of the
fragments composition, could potentially test this hypothesis.

Irrespective of its architecture, maintenance of a Dyson sphere is
extremely challenging since the structure will be punctured by a few
billion human-size asteroids every year for a planetary system that
resembles the Solar system. Over a few billion years, the structure
would have holes of a few meters in size every hundred
meters. Interestingly, the estimated size of 1I/`Oumuamua was about a
hundred meters~\citep{2019MNRAS.489.3003M}, potentially dictated by
the asteroid bombardment rate of an old Dyson sphere in a Sun-like
planetary system.

Small holes would be more abundant. The holes would cover a
significant fraction of the Dyson sphere surface from bombardment by
micrometeorites on sub-centimeter scales. This impact statistics is
currently measured empirically, as JWST is being hit by a dust-sized
particle every month. Without repair, a billion-year old Dyson sphere
would resemble a fishing net which lets a substantial fraction of the
starlight out.

In summary, once a civilization abandons its Dyson sphere this
technosignature is expected be punctured by micrometeorites and lose
its functionality on a timescale much shorter than the lifetime of the
host star. Studies of the composition of interstellar objects offers a
new way to constrain the abundance of broken Dyson spheres,
irrespective of their age and deteriorated infrared characteristics.

\bigskip
\bigskip
\section*{Acknowledgements}

This work was supported in part by Harvard's {\it Black Hole
  Initiative}, which is funded by grants from JFT and GBMF. 

\bigskip
\bigskip
\bigskip

\bibliographystyle{aasjournal}
\bibliography{m}
\label{lastpage}
\end{document}